\documentclass[useAMS,usenatbib]{mn2e}
\usepackage{graphicx,txfonts}
\pdfoutput=1

\usepackage[usenames,dvipsnames]{color}

\newcommand{\cms}{\hbox{${\rm cm\,s}^{-1}$}}
\newcommand{\ms}{\hbox{${\rm m\,s}^{-1}$}}
\newcommand{\kms}{\hbox{${\rm km\,s}^{-1}$}}
\newcommand{\SNR}{\hbox{${\rm SNR}$}}

\newcommand{\bspsmall}{\vspace{0.5cm}\small\noindent This paper has been typeset
from a \TeX/\LaTeX\ file prepared by the author.\normalsize}

\title[Frequency combs for astronomy]{Laser frequency comb techniques for precise astronomical spectroscopy}

\author[M. T. Murphy et al.]{Michael T. Murphy,$^{1}$\thanks{E-mail:
    mmurphy@swin.edu.au (MTM)} Clayton R. Locke,$^{2}$ Philip S. Light,$^{2}$ Andre N. Luiten,$^{2}$\newauthor Jon S. Lawrence$^{3,4}$\\
  $^{1}$Centre for Astrophysics and Supercomputing, Swinburne University of Technology, Hawthorn, Melbourne, Victoria 3122, Australia\\
  $^{2}$Frequency Standards and Metrology Group, School of Physics, University of Western Australia, Nedlands, WA 6009, Australia\\
  $^{3}$Australian Astronomical Observatory, Eastwood, Sydney, NSW 1710, Australia\\
  $^{4}$Department of Physics and Astronomy, Macquarie University, NSW 2109, Australia
 }

\begin{document}

\date{Accepted 2012 January 30. Received 2012 January 4; in original form 2011 December 4}

\pagerange{\pageref{firstpage}--\pageref{lastpage}}

\pubyear{2012}

\maketitle

\label{firstpage}

\begin{abstract}
  Precise astronomical spectroscopic analyses routinely assume that
  individual pixels in charge-coupled devices (CCDs) have uniform
  sensitivity to photons. Intra-pixel sensitivity (IPS) variations may
  already cause small systematic errors in, for example, studies of
  extra-solar planets via stellar radial velocities and cosmological
  variability in fundamental constants via quasar spectroscopy, but
  future experiments requiring velocity precisions approaching
  $\sim$1\,cm\,s$^{-1}$ will be more strongly affected. Laser frequency
  combs have been shown to provide highly precise wavelength
  calibration for astronomical spectrographs, but here we show that
  they can also be used to measure IPS variations in astronomical CCDs
  \emph{in situ}. We successfully tested a laser frequency comb system
  on the Ultra-High Resolution Facility spectrograph at the
  Anglo-Australian Telescope. By modelling the 2-dimensional comb
  signal recorded in a single CCD exposure, we find that the average
  IPS deviates by $<8$\,per cent if it is assumed to vary
  symmetrically about the pixel centre. We also demonstrate that
  series of comb exposures with absolutely known offsets between them
  can yield tighter constraints on symmetric IPS variations from
  $\sim$100 pixels. We discuss measurement of asymmetric IPS
  variations and absolute wavelength calibration of astronomical
  spectrographs and CCDs using frequency combs.
\end{abstract}

\begin{keywords}
  instrumentation: spectrographs -- instrumentation: detectors --
  methods: laboratory -- techniques: spectroscopic
\end{keywords}






\section{Introduction}\label{sec:intro}

A growing variety of measurements in astrophysics rely on, or are
enabled by, precise astronomical spectroscopy. For example, searches
for extra-solar planets via the radial velocity method
\citep[e.g.][]{Mayor:1995:355,Marcy:1996:L147} rely on the
reproducibility of stellar radial velocities over time-scales up to
several years. Night-to-night reproducibility better than $1\,\ms$ has
been achieved when integrated over spectral ranges of
$\sim$$1000$--$3000$-\AA\ \citep[e.g.][]{Lovis:2006:305}. Another
example is the use of relative velocity shifts between ionic and
molecular transitions in highly redshifted quasar absorption systems
to constrain cosmological variations in fundamental constants, such as
the fine-structure constant
\citep[e.g.][]{Bahcall:1967:L11,Webb:1999:884} and proton-to-electron
mass ratio
\citep[e.g.][]{Varshalovich:1993:237,Ivanchik:2005:45}. Velocity
precisions averaged over several transitions in hundreds of absorption
systems reach $\sim$$20\,\ms$
\citep{Murphy:2003:609,Malec:2010:1541,King:2011:3010,Webb:2011:191101}.

Much more precise and accurate spectroscopy would be required if the
acceleration of the Universal expansion is to be measured by
monitoring the slowly changing redshifts of Lyman-$\alpha$ forest
absorption lines in quasar spectra
\citep[e.g.][]{Sandage:1962:319,Loeb:1998:L111}. This is one possible
aim of the CODEX spectrograph proposed for the European Extremely
Large Telescope \citep[e.g.][]{Pasquini:2006:193,Liske:2008:1192}. Two
epochs of spectra are required, taken decades apart, with velocity
precision approaching 1\,\cms\ integrated over
$\sim$$1000$--$3000$\,\AA\ ranges. The calibration of individual
quasar exposures would therefore need to be reproducible to within
$\sim$1\,\cms\ over several decades. Many optical and mechanical
elements of the telescope and spectrograph may need replacing on such
a time-scale, or observations from other telescopes may need to be
employed to achieve the required signal-to-noise ratio (\SNR). Thus,
it would be preferable to \emph{absolutely} calibrate the wavelength
scale of each quasar exposure.

Laser frequency combs \citep[LFCs;
e.g.][]{Reichert:1999:59,Jones:2000:635,Udem:2002:233} are a promising
candidate for such highly precise, and possibly absolute, wavelength
calibration of astronomical spectrographs
\citep{Murphy:2007:839,Schmidt:2008:409}. The train of pulses that are
output from a frequency comb appear in frequency space as a spectrum
of lines ``modes'' uniformly spaced by the pulse repetition rate. This
repetition rate can readily be synchronised with an absolute radio
frequency reference, e.g.~a global positioning system (GPS) signal or
atomic clock. Thus, an absolute, very dense and perfectly regular
calibration spectrum could be used to calibrate astronomical
spectrographs. \citet{Murphy:2007:839} demonstrated that, in
principle, recording such a frequency comb signal on a conventional
charge coupled device (CCD) typical of modern astronomical
spectrographs would provide enough spectral information to reach the
$\sim$1\,\cms\ calibration precision required for measuring real-time
drifts in the redshifts of Lyman-$\alpha$ absorption lines in quasar
spectra.

Prototype frequency comb calibration systems are now being tested at
various observatories. The first LFC demonstration on an astronomical
spectrograph by \citet{Steinmetz:2008:1335} achieved state-of-the-art
calibration precision of $\sim$9\,\ms\ in near infra-red (IR,
1.5\,$\umu$m) astronomical spectroscopy. \citet{Wilken:2010:L16} used
a frequency-doubled Yb-doped fibre laser to calibrate a single echelle
order at $\sim$500\,nm of a highly stable vacuum spectrograph (HARPS)
with 15\,\ms\ accuracy, precision and reproducibility over
several-hour time-scales. \citet{Benedick:2010:19175} calibrated the
TRES spectrograph with an accuracy better than 1\,\ms\ at
$\sim$400\,nm using a tunable Ti:Sapphire LFC. While these tests
demonstrate the main advantage of LFC calibration -- the high density
of high-\SNR, equally-spaced modes -- some challenges discussed by
\citet{Murphy:2007:839} still remain. For example, the above tests
employed LFCs covering only small wavelength ranges (typically
$\sim$10\,nm). Simultaneously and uniformly illuminating the entire
optical wavelength range covered by modern echelle spectrographs is
highly desirable.

Another important advantage of LFC calibration is that the high
density of recorded modes and the absolute frequency scale should
enable direct characterization of many, if not all, instrumental
systematic errors \citep{Murphy:2007:839}. For example, in calibrating
a single echelle order of HARPS, \citet{Wilken:2010:L16} found
discontinuities in the calibration curve (i.e.~wavelength vs.~spectral
pixel) at regular intervals of 512 spectral pixels. They attributed
this to imperfections in CCD pixel size, shape and/or sensitivity
imparted to every 512th pixel during the manufacturing process of the
CCD mask. Even subtler problems in the CCD or spectrograph optics are
likely to limit spectroscopic accuracy unless they are first detected,
measured and then removed with LFCs.

One such problem is possible intra-pixel sensitivity (IPS)
variations. Individual CCD pixels can show large spatial variations in
sensitivity to photons depending on the manufacturing process and
depth of the depletion layer \citep{Toyozumi:2005:257}. Pixels can
also vary in size or shape and may contain small defects. Electron
counts may also be registered in a pixel when a photon enters near the
edge of a neighbouring one, depending on incidence angle. These, and
many other effects, lead to apparent IPS variations. The development
of `deep depletion' CCDs has mitigated many of these effects, but even
moderate IPS variations compared to those found by
\citet{Toyozumi:2005:257} might affect measurements of, for example,
possible variations in the fundamental constants. Thus, it is vital to
accurately measure the detailed behaviour of CCD chips -- even of
individual pixels -- if photon-limited spectroscopic precision is to
be demonstrated.

This paper reports a new demonstration of an optical LFC using a very
high resolution astronomical spectrograph with the aim of measuring
\emph{in situ} any IPS variations in a modern CCD. Section
\ref{sec:setup} explains our experimental setup. Section
\ref{sec:model} sets out the basis for our analysis to extract the IPS
variations from the comb exposures. Section \ref{sec:single} contains
our main analysis and results. Section \ref{sec:series} describes how
series of LFC exposures might allow absolute wavelength calibration of
CCDs. We summarize our main results and conclude in Section
\ref{sec:summary}.

\section{Experimental setup}\label{sec:setup}

To demonstrate that the intra-pixel sensitivity (IPS) variations in an
astronomical spectrograph's CCD can be measured \emph{in situ}, we
tested an LFC at the Anglo-Australian Telescope (AAT), Australia, on
the 9th April 2010. The Ultra-High-Resolution Facility (UHRF)
spectrograph has a very high resolution mode, with resolving power
$R\equiv \lambda/{\rm FWHM}\approx 10^6$ (where FWHM is resolution
element's full width at half maximum), which means a relatively simple
LFC can be employed (Section \ref{ssec:LFCdetails}).

\begin{figure*}
\begin{center}
\includegraphics[width=0.90\textwidth]{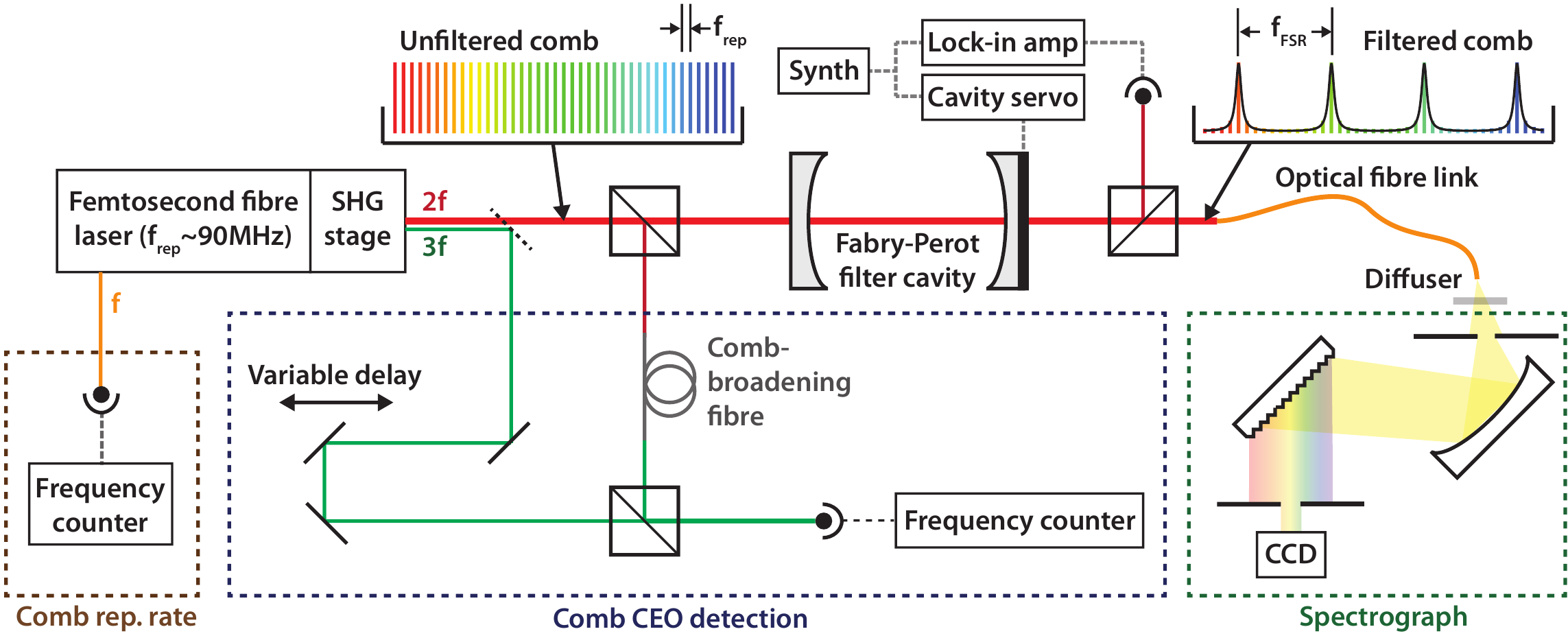}\vspace{-0.5em}
\caption{Laser frequency comb set-up with AAT/UHRF. The comb is fully
  specified by the repetition frequency, $f_{\rm rep}$, and carrier
  envelope offset (CEO) frequency, $f_{\rm CEO}$. The former is
  measured directly with a photo-diode and counter. The second
  harmonic generation (SHG) stage frequency-doubles the
  femtosecond-pulsed infrared laser (1.56\,$\umu$m) to 780\,nm (2f)
  and also produces some third harmonic light (3f) for the CEO
  measurement: the high-intensity, 780 nm light has its fairly narrow
  bandwidth broadened in a non-linear micro-structured crystal fibre
  for mixing with the 520 nm (3f) light. The frequency difference
  (`beat note') between them specifies the CEO frequency. Ten out of
  every 11 comb modes are interferometrically suppressed using a
  Fabry-Perot cavity, increasing the effective mode separation from
  94\,MHz to 1\,GHz, wide enough to be resolved from each other by the
  UHRF in its $R\approx940$\,k mode. The cavity's free spectral range
  (FSR) is locked to the filtered comb's effective repetition
  frequency by the servo electronics. The filtered comb light is
  transported to the UHRF pre-slit chamber where it is launched
  through a diffuser onto the spectrograph slit.}
\label{fig:setup}
\end{center}
\end{figure*}

Our experimental setup is illustrated in Fig.~\ref{fig:setup}. After
summarizing the basics of LFCs immediately below, we describe the
important elements of our particular setup, including our LFC, the
interface with the UHRF spectrograph and the CCD whose IPS variations
we seek to measure. We also describe the data reduction steps taken
before the analyses described in subsequent sections.

\subsection{Laser frequency comb basics}\label{ssec:LFCbasics}

A laser will generate light at discrete frequency modes throughout the
spectral bandwidth of its gain material as long as the optical gain
exceeds the losses for a full round-trip in the laser cavity at each
of the mode frequencies. The frequency of a mode is determined by
ensuring that there is a whole number of wavelengths in one cavity
round trip. Conventionally, much effort is expended in ensuring that
only a single mode can circulate at any time in the cavity; however,
for mode-locked lasers, one attempts to have many modes circulating
simultaneously. In addition, by including an intensity-dependent loss
in the cavity it is possible to bring about a mode-locked state in
which the laser's output energy is delivered in a single, intense and
temporally very short pulse. In this condition each mode maintains a
fixed and well-defined phase relationship with all other modes. The
frequency spacing between all pairs of adjacent modes, called the
repetition frequency, is exactly the same and equal to the reciprocal
of the round-trip pulse circulation time in the cavity. Further
details of the operation of mode-locked lasers can be found in
\citet{Cundiff:2003:325}.

The output spectrum of the mode-locked laser appears as a comb of modes
where the $n$th mode has a frequency, $f_n$, given by
\begin{equation}\label{eq:fn}
f_n = n f_{\rm rep} + f_{\rm CEO}\,,
\end{equation}
where $n$ is typically of the order of $10^5$. The repetition
frequency, $f_{\rm rep}$, is the mode spacing between the `teeth' of
the comb. The carrier-envelope offset frequency ($f_{\rm CEO}$)
represents the offset of each mode from an exact harmonic of the
repetition frequency, while in the time domain it expresses the rate
at which the relative phase of the carrier and pulse envelope evolves
between each pulse. For the comb to be useful as a high-precision
frequency scale, $f_{\rm rep}$ must be stabilised and $f_{\rm CEO}$
should be either stabilised or monitored. The advantage of the
mode-locked laser is that both $f_{\rm rep}$ and $f_{\rm CEO}$ are in
the radio frequency range so they can be locked (or monitored) using
relatively simple electronics. Such a comb provides a series of
regular and closely spaced modes, the absolute frequencies of which
are known to an accuracy limited only by that of the radio-frequency
reference.

\subsection{Laser frequency comb setup}\label{ssec:LFCdetails}

The first frequency combs were based on solid-state Ti:Sapphire
crystals \citep[e.g.][]{Ma:2004:1843} which, despite having
exceptional frequency stability, typically require extensive
supporting infrastructure and careful expert maintenance. In contrast,
fibre-based mode-locked lasers are low-cost, have `turn-key' operation
and, importantly for demonstration at remote astronomical
observatories, are portable. For these reasons we employ a commercial
fibre-laser for this experiment using a stabilisation scheme developed
at the University of Western Australia \citep{Locke:2009:5897}. This
laser delivers a 100-fs duration pulse at 1560\,nm with $f_{\rm
  rep}=94$\,MHz repetition frequency at an average power of approximately
250\,mW. $f_{\rm rep}$ was stabilised by detecting its 100\,th
harmonic on a very fast photodiode and referencing to a stable
microwave signal derived from a high quality synthesizer (see
Fig.~\ref{fig:setup}).

The laser output is frequency-doubled in a periodically-poled lithium
niobate (PPLN) crystal, simultaneously producing a red comb with
90\,mW at 780\,nm together with a green comb with 200\,$\umu$W at the
third harmonic, 520\,nm. The extraction of the offset frequency is as
follows (illustrated in Fig.~\ref{fig:setup}). Firstly, the green
(third harmonic) comb is separated from the red (second harmonic) comb
and travels a separate path on the optical table. Meanwhile, the red
comb is frequency-shifted to the green wavelength (520\,nm) using a
unique, highly non-linear optical fibre developed at the University of
Bath \citep{Benabid:2008:2680}. A delay arm in the beam path is
adjusted so that femtosecond pulses from both combs arrive at an
avalanche photodiode simultaneously. The difference-frequency
generated by interference of the green third harmonic comb and the
frequency-shifted red comb (both at 520\,nm) contains the offset
frequency.

Although it is possible to feed this $f_{\rm CEO}$ measurement back
into the laser, thus actively stabilising $f_{\rm CEO}$, it was not
necessary for the purposes of our LFC demonstration and IPS
measurement. This minimized the equipment required at the remote AAT
site. Because $f_{\rm CEO}$ is not multiplied by $n$ (unlike $f_{\rm
  rep}$), it was easily logged and monitored with simple electronics
during our experiments. In principle, this record could be used in
post-processing to calibrate the comb, though for the purposes of our
analysis in Section \ref{sec:single}, this was not necessary.

One important characteristic of the frequency comb used here, as with
all previous tests of LFCs on astronomical spectrographs, is that
$f_{\rm rep}$ is too low for the teeth of the comb to be resolved from
one another by the spectrograph
\citep[e.g.][]{Steinmetz:2008:1335,Wilken:2010:L16}. Indeed, this was
one challenge for the implementation of `turn-key' LFCs in astronomy
discussed by \citet{Murphy:2007:839}. Our LFC's repetition frequency,
$f_{\rm rep}=94$\,MHz, is too low even for UHRF's highest resolution
mode with $R$=940000, i.e.~FWHM=0.4\,GHz. A simple solution employed
in previous works is to filter out most modes using a Fabry-Perot (FP)
cavity. We chose the FP cavity's free spectral range (FSR) to transmit
only every 11th comb mode, leaving an effective mode separation of
1.03\,GHz -- wide enough to resolve with UHRF. In designing the
FP cavity it was firstly coupled to a continuous wave laser for end
mirror alignment to form a stable cavity. For it to operate with the
mode-locked laser a further condition must be satisfied, that is,
every 11th (in our case) pulse must map back precisely onto itself,
requiring a precision in the length of the cavity of order of a few
wavelengths. One of the mirrors on the FP cavity was set on a
fine-screw to obtain coarse cavity length tuning, and then the optical
output was monitored and used to control the cavity length via a
piezoelectric transducer (PZT) on the mirror.

The dynamic range of linear response for astronomical CCDs is limited
to $\sim$40000 photo-electrons, the CCD studied here being no
exception. Our LFC generally supplied more photons than required in
short exposures; each comb exposure only needed to be a few seconds
long. Therefore, individual exposures have very high \SNR, dominated
by the LFC photon statistics, thereby enabling constraints on the IPS
variations. We are also able to test the stability of the spectrograph
on short time-scales using a series of short exposures (see Section
\ref{ssec:series_obs}). By comparison, the usual wavelength
calibration method -- a hollow-cathode ThAr emission-line lamp --
provides only 3 emission lines near 780\,nm which provide only
moderate \SNR\ ($\sim$50\,per pixel) even after several minutes of
exposure.

Finally, we note that our LFC setup enables a simple method for
`flat-fielding' (dividing out the pixel-to-pixel sensitivity
variations in) the CCD exposures: we simply bypass the FP cavity to
provide a high-intensity spectrum whose spectral structure is
unresolved by UHRF. Again, the usual flat-field calibration method,
which utilizes a quartz lamp, would have involved averaging many
exposures lasting several minutes each.

\subsection{UHRF spectrograph setup}\label{ssec:UHRFdetails}

The UHRF is a (grating) cross-dispersed echelle spectrograph mounted
in the (east) coud\'e room of the AAT \citep{Diego:1995:323}. It was
selected for our LFC test because its very high resolution mode
($R$=940000) allows us to use a relatively inexpensive, commercial LFC
with a `low' (actually typical) $f_{\rm rep}$. The demands on the
finesse of the FP cavity mirrors (which was $F\sim300$) are also
reduced because `only' 10 out of every 11 comb modes need to be
interferometrically suppressed. And while UHRF in this mode only
allows $\sim$1\,nm of a single echelle order to be observed in a
single exposure, a wider bandwidth was not required for measuring
possible IPS variations. Also, pre-defined echelle and cross-disperser
angles were already available for observing at 780\,nm, so minimal
time was spent configuring the spectrograph itself.

The FP-filtered comb light was coupled into a single-mode optical
fibre for transport to UHRF's pre-slit chamber where it was launched
in free space through a rotating paper diffuser to over-fill the slit
in the spectral direction and almost fill it in the spatial
direction. The full UHRF resolving power of $R$=940000 was achieved by
using the smallest slit-width possible, just 40\,$\umu$m, or
0.059\,arcseconds projected on the sky. The slit had an unvignetted
length of $\sim$3.4\,mm, or $\sim$5\,arcseconds. The alignment of this
simple launch setup was checked only by eye using a television camera
system to view the 780\,nm light falling on cards placed along UHRF's
optical axis. The laser light (almost) filled the UHRF collimator
symmetrically about its centre. The alignment was not altered
throughout the exposure sequences reported in this paper.

Once the optical alignment of the launch system was finalised, the CCD
dewar was manually rotated around the optical axis and fixed so that
UHRF's spectral direction roughly aligned with the CCD rows. This
simplified the analysis by ensuring that comb spectra from individual
CCD rows spanned the entire length of the CCD.

\subsection{Charge-coupled device (CCD) details}\label{ssec:CCDdetails}

The CCD studied here, called `MITLL3' at the AAT, is chip 2 from wafer
67 (i.e.~device W76c2) of fabrication run CCID-20 from MIT/Lincoln
Labs. It has 2048$\times$4096 15$\times$15-$\umu$m pixels and was
designed to have a deep depletion layer (i.e.~high resistivity),
yielding pixels which are effectively 40\,$\umu$m thick.

Several trade-offs were considered in selecting the MITLL3 chip over
another commonly-used chip at the AAT, a back-side illuminated,
thinned EEV chip with 2048$\times$4096 13.5$\times$13.5-$\umu$m
pixels. The latter is used for precise radial velocity work at the AAT
\citep[e.g.][]{Tinney:2003:423} which might benefit from knowing the
IPS variations in that device. Nevertheless, chips from the same
fabrication run (CCID-20) as the MITLL3 are used at many telescopes
world-wide in many different instruments including, for example, the
Ultraviolet and Visual Echelle Spectrograph at the Very Large
Telescope, Chile \citep{Dekker:2000:534}. Another consideration is
that, being thinned and back-side illuminated, the EEV chip's
IPS variations may be larger and more easily detected by our LFC
test. Conversely, IPS variations in the deep-depletion MITLL3 chip
might be harder to detect.

However, the main factor in selecting the MITLL3 chip was its very low
fringing level at 780\,nm. The much larger fringing level of the EEV
would have prevented adequate flat-fielding of our comb exposures and
precluded any measurement of the IPS variations in the EEV chip.

For reference, the MITTL3 CCD pixel size in the spectral direction of
15\,$\umu$m translates to 2.63\,m\AA\ or, at 780\,nm, 129\,MHz or
101\,\ms. The nominal resolving power of $R=940000$ at 780\,nm
translates to a FWHM resolution of 8.3\,m\AA, 409\,MHz or
319\,\ms. The FWHM resolution is sampled by 3.15 pixels.

\subsection{Data reduction}\label{ssec:reduction}

\begin{figure*}
\begin{center}
\includegraphics[height=0.85\textwidth,angle=270,bb=25 17 588 773]{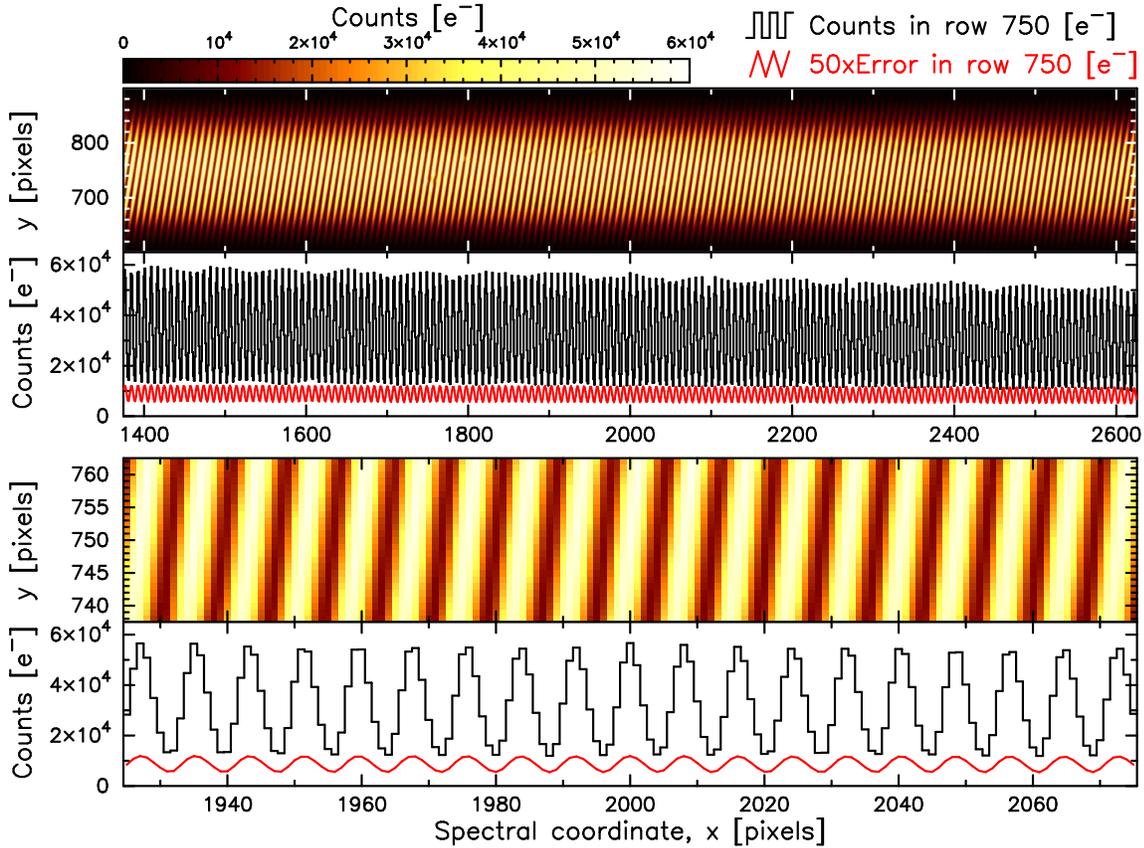}\vspace{-0.5em}
\caption{Part of a CCD exposure at 780\,nm from the AAT/UHRF frequency
  comb test (top panel) and the electron count spectrum extracted from
  a single CCD row ($y=750$; second panel from top, black
  histogram). The raw CCD image was bias-corrected and flat-fielded
  before plotting. The spatial direction is clearly tilted with
  respect to the CCD columns. The lower two panels show a zoomed-in
  portion of the upper two panels. The lowest panel shows a small
  number of the comb modes transmitted through the Fabry-Perot cavity
  (see Fig.~\ref{fig:setup}) which have a spacing of $\sim$8\,pix in
  the spectral direction ($x$ coordinate). Plotted below the
  single-row spectrum (black histogram) is the error spectrum derived
  from photon statistics (red line) exaggerated by a factor of 50.}
\label{fig:eg_2Dcomb}
\end{center}
\end{figure*}

Before any analysis of IPS variations, only basic data-reduction steps
were required. The bias level was estimated using the median of 51
zero-second CCD exposures and subtracted from each CCD exposure. As
mentioned in Section \ref{ssec:LFCdetails} above, flat-field exposures
were obtained by bypassing the FP cavity stage. A master flat-field,
formed from the median of 31 such exposures with high photo-electron
counts, was used to normalize each LFC exposure used in our
analysis. Part of a bias-corrected, flat-fielded LFC exposure is shown
in Fig.~\ref{fig:eg_2Dcomb}. Note that the spectrum's spatial
direction is tilted with respect to the CCD columns, a fact which
assists our subsequent analysis.

It is important that the error spectrum -- the flux uncertainty in
each CCD pixel -- is estimated for our analysis of possible IPS
variations. The dominant noise source is the Poisson noise of the LFC
photons; the CCD's read-noise was only $\sim$3 photo-electrons per
pixel. Thus, the error spectrum varies strongly along the spectral
direction of any given CCD row. An error image was generated for each
bias-corrected, flat-fielded LFC exposure by taking the square root of
the bias-corrected flux image and normalizing by the master flat-field
image. The error spectrum is shown for a single CCD row in
Fig.~\ref{fig:eg_2Dcomb}.

\section{Modelling the frequency comb CCD exposures}\label{sec:model}

To understand the extent of IPS variations in the CCD pixels, we model
the frequency comb flux using non-linear least squares $\chi^2$
minimisation \citep[e.g.][]{Fisher:1958}. The reliability of this
technique is demonstrated using simulated comb CCD recordings where
relevant in the following sections. It is possible that other
techniques would be effective, perhaps even simpler or faster but, for
the demonstration purposes of this paper, a full exploration of other
techniques is not important. The $\chi^2$ minimisation code is
custom-written but employs the standard Levenberg--Marquardt method
with an implementation following the basic structure of that set out
in \citet{Press:1992}.

We begin by modelling the UHRF instrumental line-shape (or
instrumental profile) from the 1-dimensional comb spectrum in
individual rows of the comb CCD exposures (Section
\ref{ssec:ILS}). That model is constrained by the assumption that
other observed spectrum properties (e.g.~average flux, frequency
dispersion per CCD pixel) vary only slowly across the CCD. This basic
model of the instrumental line-shape is then fixed (though its
parameters remain free) and the IPS variations are introduced to the
comb model in Section \ref{ssec:IPS_1D}.

\subsection{Modelling the instrumental line-shape (ILS)}\label{ssec:ILS}

For our purposes, the instrumental line-shape (ILS) is defined as the
profile formed in a single CCD row by an infinitely narrow LFC mode
after passing through the spectrograph. Note that our LFC's modes are
completely unresolved by UHRF and so each recorded comb mode will
directly represent the ILS at its observed wavelength. The frequency
stability of our LFC implies that the spectral width of each comb mode
is $\sim$1\,MHz, or $\sim$0.8\,\ms\ at the observed wavelength of
780\,nm. Given the nominal UHRF resolution of $\approx$320\,\ms, the
comb modes are very much unresolved.

For a single CCD row, we initially model the comb flux, $F_i$ (in
electron counts), in CCD pixel $i$ as the product of an `envelope
intensity' function, $E(x)$, and a `comb function', $C(x)$, both of
which are functions of position in the spectral direction of the CCD,
$x$:
\begin{equation}\label{eq:Fi}
F_i = \int_{i-1/2}^{i+1/2} E(x)\,C(x)\,dx \approx E_i \int_{i-1/2}^{i+1/2} C(x)\,dx\,.
\end{equation}

The sensitivity of $F_i$ to IPS variations is directly linked to the
fact that the comb function varies quickly on the scale of a single
CCD pixel, so it is important to integrate the model of the comb
across the pixel rather than approximate it by taking, e.g.~its value
at the pixel's centre. However, the envelope intensity of the comb,
$E(x)$ -- the variation in the measured peak intensity from mode to
mode -- varies very slowly along the spectrum; the approximation in
the right-hand-side of equation (\ref{eq:Fi}) reflects this. For all
the fits conducted in this paper, $E(x)$ is simply approximated as a 
second order polynomial.

\begin{figure*}
\begin{center}
\includegraphics[width=0.85\textwidth,bb=19 278 594 408]{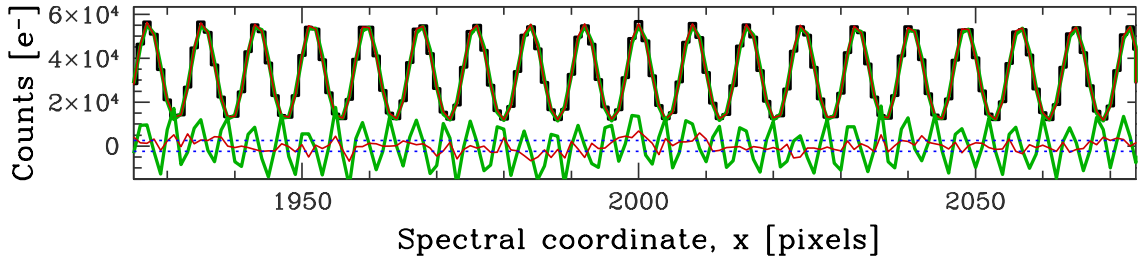}\vspace{-0.5em}
\caption{Modelling the instrumental line-shape (ILS) from small
    sections of comb spectra. The plot shows single- and
    double-Gaussian ILS models (thick green and narrow red lines,
    respectively; the latter lies almost entirely over the former)
    fitted to part of row $y=750$ of a single flat-fielded comb
    exposure (black histogram). The residuals between the two models
    and the data (again, single-Gaussian in thick green,
    double-Gaussian in narrow red), normalized to the 1-$\sigma$
    errors, are plotted below the spectrum relative to the
    $\pm$3-$\sigma$ levels (dotted blue lines). Although the two fits
    are almost indistinguishable by eye, the residual spectra show
    that the double-Gaussian ILS model is strongly preferred.}
\label{fig:fit_estILS}
\end{center}
\end{figure*}

Because the LFC modes are much narrower that the UHRF resolution, in
equation (\ref{eq:Fi}) the comb function, $C(x)$, can be written as a
sum of ILSs from each comb mode,
\begin{equation}
C(x)=\sum_{n=1}^\infty {\rm ILS}(x,x_n)\,.
\end{equation}
The $n$th mode of the comb transmitted through the Fabry-Perot cavity
occurs at pixel position $x_n$ given by
\begin{equation}\label{eq:xn}
x_n = x_0 + \omega_r n + \omega_2n^2 + \ldots\,,
\end{equation}
with $\omega_r$ the `pixel-space repetition frequency'. Equation
(\ref{eq:xn}) is analogous to equation (\ref{eq:fn}) but, because we
model the comb in pixel space rather than frequency space,
higher-order terms are required, characterised by $\omega_2$,
$\omega_3$ etc.\,. In this work we found no need to include terms
beyond second order in equation (\ref{eq:xn}) because we model such
short frequency ranges of comb spectra. Visual inspection of
Fig.~\ref{fig:eg_2Dcomb} reveals that comb modes fall on approximately
every 8th pixel -- i.e.~$\omega_r\approx8$\,pix -- without any obvious
non-linearity, i.e.~$n\,\omega_2\ll\omega_r$. Indeed, in our detailed
modelling below, $\omega_2$ is typically $\sim -1\times10^{-4}$ with
$-50\la n\la 50$.

Visual inspection of the example comb spectrum recorded in a single
CCD row (see Fig.~\ref{fig:eg_2Dcomb}) reveals UHRF's ILS to be
slightly asymmetric. Our initial guess for the ILS was therefore a
double-Gaussian model: a central Gaussian flanked by a weaker second
one with the same width. That is,
\begin{equation}\label{eq:ILSgauss}
{\rm ILS}(x,x_n) = \sum_{m=1}^M A_m\,\exp{\left[-(x-\Delta_m-x_n)^2/2\sigma_{\rm ILS}^2\right]}
\end{equation}
with $M=2$, $\Delta_1\equiv0$ and the ratio of Gaussian amplitudes,
$A_2/A_1$, normalized so that $\int_{-\infty}^\infty {\rm ILS}(x,x_n)\,dx = 1$.

Figure \ref{fig:fit_estILS} shows the model specified in equations
(\ref{eq:Fi}--\ref{eq:ILSgauss}) fitted to a small section of a single
row of one comb exposure. Also plotted is the same model but with a
single Gaussian ILS instead of a double-Gaussian (i.e.~$M=1$ instead
of 2). The two models are not readily distinguishable by eye. However,
the residual spectra clearly show that the double-Gaussian model
provides a closer match to the data. This is also reflected in the
values of $\chi^2$ per degree of freedom, $\nu$, for the two fits:
$\chi^2_\nu=29.5$ and 2.7 for the single and double-Gaussian models
respectively. We discuss why the latter value is noticeably greater
than unity in Section \ref{ssec:results_indiv}. The residual spectrum
for the double-Gaussian model, also plotted in
Fig.~\ref{fig:fit_estILS}, does not appear to show any features which
repeat on the scale of the comb mode structure, unlike that for the
single-Gaussian model. We attempted to include a third Gaussian
(i.e.~$M=3$) in the ILS but its best-fitted amplitude was typically
very small ($A_3/A_1 \la 0.02$) and, upon fitting different CCD rows
and using different CCD exposures, its separation from the main
Gaussian component was sporadic and very weakly constrained. This is
consistent with the lack of periodic structure in the residuals from
the double-Gaussian model in Fig.~\ref{fig:fit_estILS}. Also,
including the third Gaussian barely reduced $\chi^2_\nu$ in most
cases; when fitting some CCD rows from some exposures it actually
increased. That is, there is no clear evidence for additional
structure in the ILS beyond the double-Gaussian model.

\citet{Diego:1995:323} measured the UHRF ILS using a He-Ne laser. From
their figure 10, their measured ILS can be described as comprising two
main Gaussian-like components separated by $\sim$0.25\,\kms\ with
similar widths of ${\rm FWHM}\approx0.28$\,\kms\ and with one being
$\approx$20\,per cent the height of the other. This accords well with the
ILS model derived from the comb spectra: in the simple fits above, the
Gaussian components were separated by $\approx$0.23\,\kms, they had
${\rm FWHM}\approx0.30$\,\kms\ and the weaker one was $20$\,per cent
the height of the stronger one. Thus, the double-Gaussian model of the
UHRF ILS seems well justified, both from the comb spectra themselves
and the He-Ne spectra of \citet{Diego:1995:323}.

When fitting different parts of the same CCD exposure studied in
Figs.~\ref{fig:eg_2Dcomb} \& \ref{fig:fit_estILS}, and the others
taken over the course of our LFC test, we found no significant
variations in the ILS structure. That is, no further Gaussians seem to
be required. Nor do we observe large variations in the parameters of
the two Gaussians constituting our preferred ILS model; when modelling
a 2-dimensional comb spectrum in Section \ref{sec:single} we
explicitly test for this possibility. However, throughout this rest of
this work we adopt the basic double-Gaussian model of the ILS with
free parameters as specified in equation (\ref{eq:ILSgauss}).

Finally, a by-product of the above fits is the statistical uncertainty
on $x_0$ in equation (\ref{eq:xn}), the zero-point of the comb
function. For example, the double-Gaussian fit to the 150 spectral
pixels in Fig.~\ref{fig:fit_estILS} returns a 1-$\sigma$ uncertainty
on $x_0$ of $9.9\times10^{-3}$\,pix, or $\approx$1\,\ms. In the
absence of intra-pixel sensitivity variations (which we seek to
measure in this work), this would correspond to the precision with
which such short sections of the comb spectrum could be
wavelength-calibrated.

\subsection{Modelling CCD intra-pixel sensitivity (IPS) variations}\label{ssec:IPS_1D}

In our comb test with UHRF we sought to detect or limit any
IPS variations only in the spectral direction across the CCD
pixels. This is possible because the flux measured in a given pixel
will depend on the IPS variations most strongly when the comb flux
changes substantially across a pixel. From the example comb exposure
shown in Fig.~\ref{fig:eg_2Dcomb} it is clear that the flux variations
in the spatial direction are much more slowly varying compared with
those in the spectral direction. Therefore, we do not attempt to model
or detect IPS variations in the spatial direction across pixels.

Thus, when fitting the comb flux in a single CCD row, or part thereof,
we insert a term for the IPS map, ${\rm IPS}_i(x)$, in equation
(\ref{eq:Fi}) to obtain
\begin{equation}\label{eq:Fi_IPS}
F_i \approx E_i \int_{i-1/2}^{i+1/2} C(x)\,{\rm IPS}(x)\,dx\,,
\end{equation}
where, as discussed above, ${\rm IPS}(x)$ is the same for all $i$, or
${\rm IPS}_i(x)={\rm IPS}(x)$.

Our aim here is to measure the average IPS variations, not the IPS map
for individual pixels. The latter possibility is discussed in Section
\ref{sec:series}. We therefore make two important assumptions in the
modelling that follows:
\begin{enumerate}
\item The IPS variations in all CCD pixels under consideration are
  assumed to be the same. While previous laboratory studies have shown
  that IPS variations are different in different pixels, they also
  demonstrate a high degree of correlation across large portions of
  the CCD \citep[e.g.][]{Toyozumi:2005:257}.
\item The IPS variations are modelled as a symmetric function about
  the centre of the pixel (in the spectral direction). We report
  results here from a Gaussian function with variable width, truncated
  at the pixel edges and a base fixed at zero with no slope. For
  Gaussian widths larger than ${\rm FWHM}\sim0.5$ pixels, which we
  find are always preferred for the UHRF CCD, a variable base would be
  closely degenerate with the width parameter. Similar functions --
  e.g.~symmetric linear slopes away from a central IPS peak (or
  trough) to the pixel edges -- give very similar results. It is also
  important to realise that a variable baseline slope would be closely
  degenerate with the parameters in equation (\ref{eq:xn}) which
  characterize the position of the comb modes. That is, a general
  slope in the IPS map of the fitted pixels is not detectable in our
  work. We discuss this further in Section \ref{ssec:series_sim}.
\end{enumerate}

We will see in Section \ref{ssec:results_indiv} that it is only by
fitting many hundreds of pixels (all assumed to have the same IPS map)
that the degeneracies in our fitted model are reduced to a level where
the Gaussian width of the IPS model can be constrained. For example,
it is the small asymmetry in the UHRF ILS which allows us to constrain
a symmetric IPS map, as we have chosen to use here. Nevertheless, this
proves sufficient for measuring IPS variations in the MITLL3 chip
using the assumptions above. As we demonstrate in Section
\ref{sec:series}, it may be possible to measure non-symmetric average
IPS maps in future using additional information.

To summarize the above discussion, we may write our model of the IPS
map for all pixels as a function of $x$ (where we note that $x=0$ is
the middle of a pixel) as
\begin{equation}\label{eq:IPS}
{\rm IPS}_i(x) = {\rm IPS}(x) = B\,\exp{\left(-x^2/2\sigma_{\rm IPS}^2\right)}\,,
\end{equation}
where the amplitude, $B$, is normalized such that
$\int_{-1/2}^{1/2}{\rm IPS}(x)\,dx=1$. Here, $\sigma_{\rm IPS}$,
measured in units of pixels, is the IPS Gaussian width parameter which
we measure. For example, $\sigma_{\rm IPS}\gg 1$\,pix indicates a flat
average IPS map whereas $\sigma_{\rm IPS}\la 1$\,pix indicates a
strongly variable average IPS map.

Finally, note that the form of equation (\ref{eq:Fi_IPS}) is
convenient because the integrand is the sum of a product of Gaussian
functions, thereby allowing trivial numerical integration using the
well-known Gamma function \citep[e.g.][]{Press:1992}.

\section{Average intra-pixel sensitivity map from a single comb
  exposure}\label{sec:single}

To measure the IPS variations using a single CCD exposure we assume
that they are the same in all pixels and utilise the fact that
each pixel (i.e.~``copy'' of the IPS map) is probed with a different
flux distribution. By using a well-constrained model of the flux
distribution across many pixels, $\sigma_{\rm IPS}$ -- the Gaussian
width of the average intra-pixel sensitivity (IPS) map in our model --
can be measured. From preliminary fits to $\sim$200-pixel sections of
single CCD rows, it became clear that $\sigma_{\rm IPS}$ was covariant
(i.e.~somewhat degenerate) with several other parameters in our comb
and instrumental line-shape models, e.g.~$\sigma_{\rm ILS}$, the width
of the two Gaussian ILS components. Rather than attempt to fit many
more pixels in a single CCD row, which may necessitate higher-order
representations of the intensity envelope function, $E(x)$, and comb
function, $C(x)$, we instead utilized the full 2-dimensional (2D) comb
information in our UHRF CCD exposures.

\subsection{2D modelling of a single comb exposure}

The example CCD exposure in Fig.~\ref{fig:eg_2Dcomb} illustrates how
the comb signal occupies more than 100 pixels in the spatial
direction. That is, there are many ``copies'' of almost the same comb
signal available for constraining $\sigma_{\rm IPS}$. This means that,
under our assumption that all CCD pixels have the same IPS variations,
we can fit many CCD rows simultaneously by adding a very small number
of additional free parameters to the model in equation
(\ref{eq:Fi_IPS}).

Firstly, each of the terms of the comb function in equation
(\ref{eq:xn}) now becomes a function of CCD row, $j$. That is, the
$n$th mode in row $j$ can be written as
\begin{equation}
x_n^j = x_0(1+a_1j+a_2j^2) + \omega_rn(1+b_1j) + \omega_2n^2
\end{equation}
where we have now explicitly truncated the series at second order in
$n$ and limited the order in $j$ in each term.  Figure
\ref{fig:eg_2Dcomb} clearly shows that the spatial direction of the
spectrograph is tilted with respect to the CCD columns, i.e.~$a_1>0$
to account for this tilt. But after exploring many 2D fits to
various comb exposures and portions of the CCD, we found no evidence
for distortions in this tilt, or for variations in the pixel-space
repetition frequency with row. That is, both $a_2$ and $b_1$
  were consistent with zero. Therefore, for all fits reported here we
fixed $a_2$ and $b_1$ to zero.

Similarly, each of the parameters of the instrumental line-shape (ILS)
could vary with $j$. By appropriately altering equation
(\ref{eq:ILSgauss}) to include polynomial dependence on $j$ for all free
parameters (e.g.~$\sigma_{\rm ILS}$), we explored various fits to
search for evidence of this. However, we found no significant
variations in the ILS in the spatial direction.

Finally, note that while the intensity envelope function, $E_(x)$,
clearly varies (slowly) with CCD row, $j$, it will not be strongly
covariant with any IPS variations, the measurement of which is our
main aim. Therefore, we simply allowed the parameters of $E(x)$ to be
independently determined for each $j$.

\subsection{Main results: limits on IPS variations from individual
  exposures}\label{ssec:results_indiv}

We searched for IPS variations in our 10 highest SNR (20-s long)
exposures. In each exposure we concentrated on fitting a
50$\times$320\,pix (rows\,$\times$\,columns) region near the centre
(highest SNR) of the comb signal. This represents about half the
comb's spatial extent and about 10\,per cent of its spectral extent on
the CCD. Preliminary fits showed that some portions of the fitted area
were not well flat-fielded. From inspection of the raw images and the
master flat-field frame, we determined that dust particles on the UHRF
optics were most likely resulting in small flat-field residuals which
varied slightly from exposure to exposure or at least between the comb
and flat-field exposures. The portions of the fitted area were the
same in the 10 exposures we studied in detail. Therefore, we simply
masked these portions out of our $\chi^2$ minimization analysis. We
note in passing that 2D fits to different regions of the CCD did not
reveal any obvious periodic CCD defects like those identified by
\citet{Wilken:2010:L16}.

\begin{figure*}
\begin{center}
\includegraphics[width=0.85\textwidth,bb=50 371 594 714]{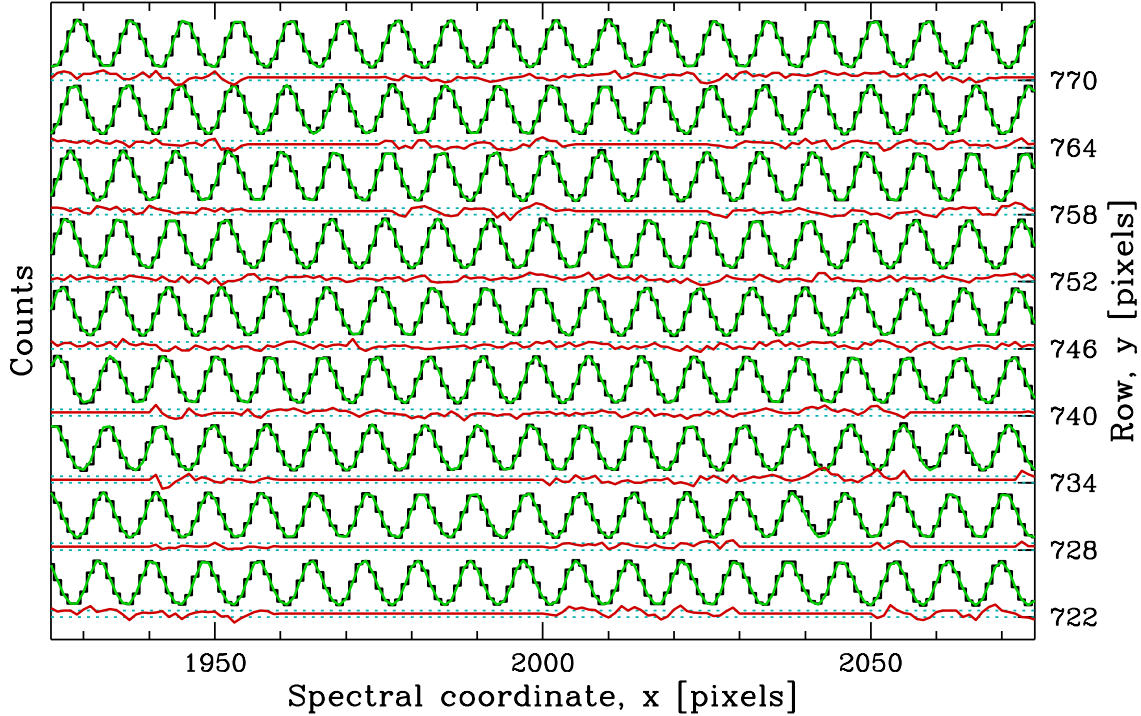}\vspace{-0.5em}
\caption{Part of a 2D fit to a single comb exposure. The flat-fielded
  spectra from every 6th fitted row are shown as black histograms
  overlaid by the model (green line) which was fitted to 50 rows and
  320 pixels in the spectral direction. The normalized residuals are
  plotted below the spectrum of each row relative to the
  $\pm$3-$\sigma$ levels (dotted blue lines).  Note that some sections
  of residuals are all zero; these parts of the CCD have been masked
  due to problems with flat-fielding (see text). Note that the tilt of
  the spatial direction across the CCD columns is evident.}
\label{fig:fit_2d}
\end{center}
\end{figure*}

Figure \ref{fig:fit_2d} shows part of a 2D fit to a representative
exposure. In general, the fit is clearly a reasonable representation
of the recorded data. The $\chi^2$ per degree of freedom, $\nu$, was
similar for all 10 exposures, $\chi^2_\nu\approx2.9$ which, while not
$\approx$1, is not unreasonable, especially in the possible presence
of additional, low-level flat-fielding errors. The residual spectra
generally do not show coherent structures which repeat over $\sim$8
pixel ranges, i.e.~the transmitted comb mode spacing. Figure
\ref{fig:fit_2d} clearly shows some exceptions to this general
statement. However, we did not observe any obvious correlation between
the positions of these exceptions in different exposures. They may
very well result from IPS variations in small clusters of pixels, but
we cannot quantitatively address that possibility with the data in
hand. Again, we attempt here to constrain a model of the average IPS
variations for the whole fitted area.

For all of the 10 individual exposures, simply including $\sigma_{\rm
  IPS}$ as a free parameter in the fit yielded best-fit values
$\sigma_{\rm IPS}>2$\,pix in all cases. This was independent of the
starting guess for $\sigma_{\rm IPS}$ (and other parameters in the
model), which was generally $\sim$0.5\,pix. In most cases $\sigma_{\rm
  IPS}$ grew from the starting guess value to $100$\,pix, the
arbitrary limit imposed in our code. And in all cases, the best fit
value was entirely consistent with $\sigma_{\rm IPS}=\infty$. That is,
under the assumptions of our IPS model, we do not detect significant
average IPS variations.

\begin{figure}
\begin{center}
\includegraphics[width=0.95\columnwidth,bb=18 269 584 712]{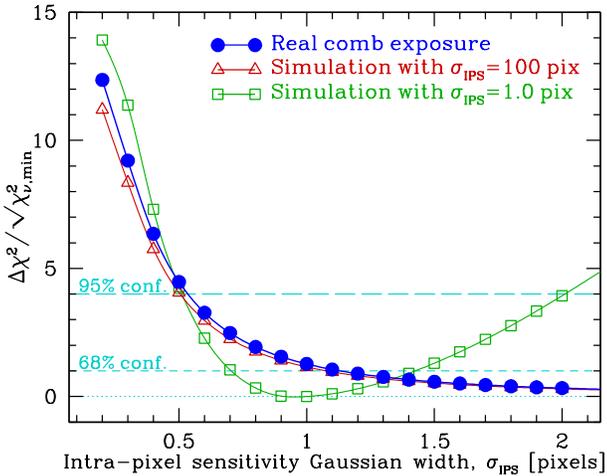}\vspace{-0.5em}
\caption{Results from 2D fit to the central part (50$\times$320
  pixels) of a single flat-fielded comb exposure. The plot represents
  the increase in $\chi^2$ from its minimum value, $\chi^2_{\rm min}$
  (i.e.~$\Delta\chi^2\equiv\chi^2-\chi^2_{\rm min}$) as a function of
  the Gaussian width of the intra-pixel sensitivity (IPS) variations,
  $\sigma_{\rm IPS}$. $\Delta\chi^2$ is normalized by
  $\sqrt{\chi^2_{\nu,{\rm min}}}$ where $\chi^2_{\nu,{\rm min}}$ is
  the minimum value of $\chi^2$ per degree of freedom, $\chi^2_\nu$,
  so that confidence intervals can be calculated directly from the
  plot. $\sigma_{\rm IPS}$ is unbounded above in the fit to the real
  comb exposure (blue curve with circular points). That is, we cannot
  rule out the null hypothesis of zero average IPS variations in that
  exposure. However, $\sigma_{\rm IPS}\la0.5$\,pix is ruled out at
  95\,per cent confidence ($\Delta\chi^2/\sqrt{\chi^2_{\nu,{\rm
        min}}}\approx4$). A fit to a simulated version of the same comb
  exposure with a flat IPS map inserted yield very similar results
  (red curve with triangular points). The green curve with square
  points shows the results from a fit to the same simulation in which
  a non-flat IPS map was used, i.e.~$\sigma_{\rm IPS}=1$\,pix. The
  definite minimum demonstrates that non-flat IPS variations could
  have been detected given the assumptions in our model of the comb
  exposures.}
\label{fig:IPSresults}
\end{center}
\end{figure}

To ensure that our $\chi^2$ minimization code was functioning
correctly we fitted simulated versions of the 10 comb spectra. The
real comb flux was replaced with the best-fit model (without IPS
variations) with Gaussian noise added which was derived using the real
flux error estimates. The synthetic realisations were fitted in the
same way, and with the same free parameters, as the real
spectra. Figure \ref{fig:IPSresults} illustrates one aspect of the
comparison between the results from one comb exposure and the average
results from 25 simulated versions of it. By holding $\sigma_{\rm
  IPS}$ fixed at certain values and plotting the deviation in $\chi^2$
from its minimum, $\Delta\chi^2$, we gain a more accurate
understanding of the minimum value of $\sigma_{\rm IPS}$ allowed for
each exposure. Because $\chi^2_\nu$ is not close to unity for the real
exposures, but it is (by construction) for the simulations, we can
test the veracity of the usual assumption that confidence intervals
for $\sigma_{\rm IPS}$ may be derived after normalizing $\chi^2$ by
$(\chi^2_{\nu,{\rm min}})^{1/2}$, the square root of the minimum
$\chi^2_\nu$. Under this normalization, Fig.~\ref{fig:IPSresults}
shows very similar results for the real and simulated spectra.

For the particular exposure studied in Fig.~\ref{fig:IPSresults} we
can rule out $\sigma_{\rm IPS}<1.0$\,pix ($<0.5$\,pix) at the 68\,per
cent (95\,per cent) significance level. These constraints on
$\sigma_{\rm IPS}$ were uniformly replicated for the same fitted
region across the 10 exposures we studied in detail. By co-adding the
plots of $\Delta\chi^2/(\chi^2_{\nu,{\rm min}})^{1/2}$ versus
$\sigma_{\rm IPS}$ for all 10 exposures, and recalling that we fitted
the same region of the CCD in all cases, we form a joint constraint on
the Gaussian width of our assumed IPS variation map: $\sigma_{\rm
  IPS}>1.2$\,pix at 99.7\,per cent (3-$\sigma$) confidence. This value
of $\sigma_{\rm IPS}$ corresponds to variations in the IPS across
the spectral direction of the pixel of $<8$\,per cent.

Instead of a decrease in the IPS from the pixel centre to its edges,
we also explored an IPS map with a minimum at the pixel centre by
adding unity to equation (\ref{eq:IPS}) and allowing $B<1$. We find
very similar results to those above: $\sigma_{\rm IPS}$ is constrained
to be $>1.3$\,pix at 99.7\,per cent confidence. That is, we rule out
IPS variations of $>7$\,per cent at 3-$\sigma$ significance under the
assumptions of our modified (up-side down Gaussian) model.

Finally, to demonstrate that detecting non-zero IPS variations is
possible using our $\chi^2$ minimization technique (under the
assumptions of our 2D modelling), we again fit simulated versions of
the real comb spectra, this time with finite values of $\sigma_{\rm
  IPS}$. The SNR of the simulated spectra again matched the
corresponding real spectra. In all simulations the best-fit value of
$\sigma_{\rm IPS}$ broadly matched the input value. This was simple to
assess for input values $\sigma_{\rm IPS}\la1.5$\,pix but judging
formal statistical consistency between the input and best-fit values
was more difficult when $\sigma_{\rm IPS}$ was larger. The ultimate
reason for this is that the IPS map is very flat for large
$\sigma_{\rm IPS}$. This is illustrated in Fig.~\ref{fig:IPSresults}
which shows the variation in $\chi^2$ with fixed fitted values of
$\sigma_{\rm IPS}$ for a single realisation with an input $\sigma_{\rm
  IPS}=1$. In this example there is a clear minimum in $\chi^2$ near
the input value of $\sigma_{\rm IPS}$. However, it is also clear that
$\chi^2$'s sensitivity to $\sigma_{\rm IPS}$ diminishes with
increasing $\sigma_{\rm IPS}$, i.e.~the $\chi^2$ curve is not
parabolic at $\sigma_{\rm IPS}\ga1$. Thus, the formal statistical
error on $\sigma_{\rm IPS}$ derived from the covariance matrix in our
$\chi^2$ minimization scheme will not be reliable for large
$\sigma_{\rm IPS}$. This problem clearly also applies to the
constraints on $\sigma_{\rm IPS}$ derived from our real comb exposures
above. This is the reason we used the $\chi^2$ curve for the fit to
each exposure, rather than the formal error from the covariance
matrix, to derive confidence limits on any IPS variations.

To summarize the results above, we have demonstrated that any
symmetric component of the average intra-pixel sensitivity (IPS)
variations can be detected in a single LFC exposure provided that many
hundreds of pixels can be modelled simultaneously. The AAT/MITLL3
chip's average IPS map deviates from being flat by $<8$\,per cent in a
symmetric manner about the pixel centre in the direction along CCD
rows. However, under our model assumptions, we cannot measure or rule
out a significant asymmetric component to the average IPS map. For
example, the IPS map may slope from one side of the pixel to the
other, but we cannot detect this by modelling a single exposure. The
simple reason for this is that such a slope is closely degenerate with
the measured position of the comb with respect to the CCD pixel
grid. We return to this point in Section
  \ref{ssec:series_sim}. In the next Section we discuss how a series
of offset comb exposures might allow that degeneracy to be removed.

\section{Intra-pixel sensitivity map from a series of offset comb
  exposures}\label{sec:series}

\subsection{Observed series}\label{ssec:series_obs}

With a view to detecting any very small average IPS variations, the
10$\times$20-s comb exposures with high \SNR\ utilized in the previous
section were taken in series with small offsets in the LFC signal
applied between exposures. The offsets were generated by slightly
altering the repetition frequency of the mode-locked laser, $f_{\rm
  rep}$, which was observed to be stable between exposures. The offset
in $f_{\rm rep}$ between each exposure was $\sim$15\,Hz so that the
comb modes at 780\,nm would shift by $\sim$50\,\ms, or 0.5 spectral
pixels. The carrier-envelope offset frequency, $f_{\rm CEO}$, was also
monitored and fluctuations were negligible ($<10$\,\ms) during the
series of exposures.

\begin{figure}
\begin{center}
\includegraphics[width=0.95\columnwidth,bb=17 269 580 583]{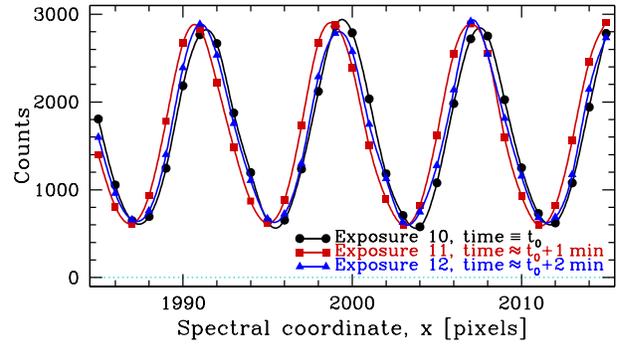}\vspace{-0.5em}
\caption{Shifts between comb spectra on short time-scales. Plotted are
  small sections of a single row of three 1-s long flat-fielded comb
  spectra (traced by spline curves to between distinguish them). The
  exposures were taken approximately 1\,min apart and all three had
  the same repetition frequency ($f_{\rm rep}$) and carrier envelope offset
  frequency ($f_{\rm ceo}$), yet they show clear, stochastic shifts
  of $\sim$0.4 pixels.}
\label{fig:shifts}
\end{center}
\end{figure}

However, upon inspecting the series of offset comb exposures, it was
apparent that the observed shifts in the comb modes did not accurately
reflect those expected from the offsets applied to $f_{\rm rep}$
between successive exposures. To investigate this we took another
series of 10 short (1-s) exposures with no offsets in $f_{\rm rep}$
applied between them. That is, including the CCD read-out time of each
exposure, a 1-s exposure was taken approximately every minute and,
since neither $f_{\rm rep}$ nor $f_{\rm CEO}$ changed appreciably
during the 10\,min series, we expected no shifts between the comb
modes. However, we discovered fairly large stochastic shifts between
successive comb exposures. These were typically $\la$0.2 pixels but
some were as large as $\sim$0.4 pixels. An example from 3 successive
1-s exposures is shown in Fig.~\ref{fig:shifts}. Furthermore, no two
successive exposures had zero shift between them, indicating that the
time-scale for these shifts is substantially less than 1\,min.

The origin of these shifts is not clear from our comb spectra, but the
fast time-scale suggests that a purely thermal drift within UHRF is an
unlikely culprit and that a more likely explanation is small
mechanical oscillations or jumps/relaxations in the
spectrograph. UHRF's main components and camera are mounted on a large
($\sim$3-m), vibration-damped and isolated bench. The slit assembly is
mounted on the same bench. However, the pre-slit area upon which our
LFC launch setup was mounted (see Fig.~\ref{fig:setup}) is, at least
in principle, decoupled from the spectrograph bench. Thus,
oscillations induced by moving components in the pre-slit area
(e.g.~the rotating diffuser) should not cause the slit itself to
oscillate. However, it remains possible the other equipment in the
pre-slit area might transmit some vibrations to the slit assembly,
which may have been responsible for the short time-scale, stochastic
shifts we observe. If such oscillations/vibrations did not originate
from our equipment, they may also affect astronomical observations
with UHRF; however, they are unlikely to have been detectable in
UHRF before because of the much longer ThAr exposure times required
(i.e.~several minutes). Observations with the University College
London Echelle Spectrograph, UCLES, which uses the same coud\'e room
and slit assembly as UHRF, might also be affected. We would be
interested to learn whether users of UCLES have identified such
problems before, but to our knowledge they have not been reported in
the literature.

Unfortunately, these stochastic shifts between exposures mean that we
cannot reliably model the observed series of offset comb exposures to
constrain IPS variations. One should consider whether the analysis of
the individual 20-s exposures in Section \ref{sec:single} was affected
by this instability. The effect of quickly varying shifts in the comb
position, averaged over a comparably long exposure, should be to
increase the observed comb mode width, or $\sigma_{\rm ILS}$ as
parametrized in our comb models. We do not observe a larger
$\sigma_{\rm ILS}$ for the 10$\times$20-s comb exposures compared to
the 10$\times$1-s ones: $\left<\sigma_{\rm ILS}\right>=1.632$ compared
to $\left<\sigma_{\rm ILS}\right>=1.629$\,pix, respectively. However,
the root-mean-square (RMS) variations in these quantities -- 0.006 and
0.011\,pix, respectively -- may be too large to detect the increase in
width expected: if random comb shifts have ${\rm RMS}\sim0.2$\,pix
then $\sigma_{\rm ILS}$ would broaden by just $\sim$0.01\,pix. Even if
this effect were present in the 20-s comb exposures, our model of the
comb flux across the CCD pixels is still likely to be adequate and so
the constraints on IPS variations we derive are unlikely to be
strongly affected.

\subsection{Simulated series}\label{ssec:series_sim}

While the short time-scale instabilities in UHRF prevent us from
constraining the average IPS variations using a real series of offset
comb exposures, we can nevertheless demonstrate this possibility using
simulated comb spectra. We simulated 10 spectra, each with a single
CCD row 320 pixels long. The comb signal was generated using the
best-fitting parameters derived from row 750 of the CCD exposure
studied in Section \ref{ssec:results_indiv} (see
Figs. \ref{fig:eg_2Dcomb} \& \ref{fig:fit_estILS}), including the
double-Gaussian instrumental line-shape. Gaussian random noise was
added to each spectrum independently, with the average \SNR\ set to
the same value as in the real CCD exposure. However, we introduced
0.8-pixel offsets between the comb functions for successive
spectra. We then simultaneously fitted comb functions to all 10
spectra, the key assumption being that the relative offsets between
spectra were known; if the LFC system is absolutely calibrated then
this will be true to very high precision.

\begin{figure}
\begin{center}
\includegraphics[width=0.95\columnwidth,bb=44 181 580 636]{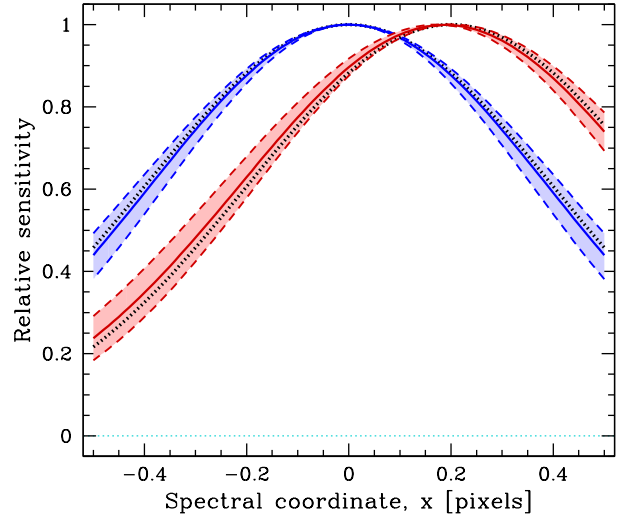}\vspace{-0.5em}
\caption{Recovery of average intra-pixel sensitivity (IPS) map from
  series of offset comb spectra. Ten single-row, 320-pixel comb
  spectra were simulated with the same parameters (similar to those of
  our real comb exposures), but with offsets of 0.8\,pixels between
  their respective comb functions and different noise
  realisations. The blue (symmetric about $x=0$) shaded region shows
  the 1-$\sigma$ uncertainty range in the measured IPS map (solid
  curve) when only the Gaussian width of the IPS model is a fitted
  free parameter. The red (asymmetric) region and curves represent the
  results when both the Gaussian IPS width and its position along $x$
  are free parameters and where it is assumed that the absolute
  position of the comb with respect to the CCD pixel grid is known. In
  both cases the input IPS model is shown as a black dotted curve.}
\label{fig:simIPS}
\end{center}
\end{figure}

Figure \ref{fig:simIPS} shows the results of the fits to the simulated
spectra. Similar to the simulations in the previous section, the input
IPS model was a Gaussian with $\sigma_{\rm IPS}=0.4$\,pix centred at
$x=0$. That is, the Gaussian was centred at the middle of the pixel
and $\sigma_{\rm IPS}$ was a free parameter. Note that the ILS and
envelope function parameters were also free in this fit. Figure
\ref{fig:simIPS} shows that this symmetric IPS model is recovered with
a relatively small uncertainty in $\sigma_{\rm IPS}$ of
0.03\,pix. This demonstrates that a series of offset comb spectra can
be used to characterise the symmetric component of the average IPS
map.

The main reason a series of offset spectra can strongly constrain the
IPS variations, even with fairly few pixels (e.g.~a single CCD row)
assumed to have the same IPS map, is that the comb flux distribution
across each pixel is changed by a known amount when the comb is offset
by a known amount. However, as with the 2D analysis of a single
exposure in Section \ref{sec:single}, any asymmetric component of the
IPS map is highly degenerate with the overall position of the
comb. That is, without knowing where the comb modes fall with respect
to the CCD pixel grid, asymmetries in the average IPS map cannot be
reliably determined. If some additional information could be gained
about where the comb light is striking pixels, perhaps by adjusting a
physical mask in front of the CCD, then the same fitting technique
could be used to recover asymmetries in the IPS map.

As an example of this, we simulated another ten 320-pixel, single-row
comb spectra in the same way as above but with an asymmetric input IPS
map, i.e.~a Gaussian, as before, but shifted with respect to the pixel
centre. It was then assumed that the absolute position of the comb
modes with respect to the pixel grid was known. Figure
\ref{fig:simIPS} shows that the fitting technique indeed recovers the
asymmetry in the input IPS model with high accuracy. Much higher
precision measurements of the average IPS variations should be
possible if the instrumental line-shape (ILS) is also known accurately
\emph{a priori} (or measured independently) and not simultaneously
constrained by the comb spectra. With a large enough number of
high-\SNR\ exposures with very small, known offsets, and with known
comb-mode--CCD-grid positioning and ILS information, it should be
possible to measure the IPS map of every illuminated CCD pixel.

\section{Summary}\label{sec:summary}

We reported the successful test of a laser frequency comb (LFC) system
on the Ultra High Resolution Facility (UHRF) spectrograph at the
Anglo-Australian Telescope (AAT). Figure \ref{fig:setup} shows the
experimental setup used. The high spectral resolution of UHRF
($R\approx940000$) allowed a commercial fibre laser with a fairly
typical repetition frequency ($f_{\rm rep}$) to be used.

A Fabry-Perot cavity was used to widen the effective separation
between modes to $\sim$1GHz (by suppressing 10 of every 11 modes) so
they could be resolved apart by UHRF. The LFC spectra were recorded
with the `MITLL3' CCD chip at $\approx$780\,nm in a single echelle
order of UHRF covering $\approx$10\,\AA. An example exposure is shown
in Fig.~\ref{fig:eg_2Dcomb}.

Our main goal in the LFC test was to demonstrate that LFC spectra
could be used to characterize the average intra-pixel sensitivity
(IPS) map of astronomical CCDs \emph{in situ}. This meant that
actively self-calibrating the LFC (i.e.~controlling $f_{\rm rep}$ and
$f_{\rm CEO}$, not just measuring/monitoring them) was not necessary,
which minimized the equipment required on-site.

We demonstrated the measurement of IPS variations using a single comb
exposure by modelling the comb flux in 50$\times$320 pixels in the
spatial and spectral directions, respectively -- see
Figs.~\ref{fig:fit_2d} \& \ref{fig:IPSresults}. For a simple Gaussian
model of the average IPS variations where the sensitivity falls off
symmetrically about the pixel centre, the IPS map of the AAT/MITLL3
chip deviates from flat by $<8$\,per cent. A similar constraint on an
inverted map (i.e.~sensitivity increasing towards the pixel edges) was
also obtained.

The instrumental line-shape (ILS) was also modelled in this process,
and we found that a double-Gaussian ILS was sufficient to fit our comb
spectra -- see Fig.~\ref{fig:fit_estILS}. This ILS is very similar to
that found when commissioning UHRF \citep{Diego:1995:323}.

By slightly adjusting the repetition frequency between comb exposures (by
changing the length of the fibre laser's free-space section), we
obtained a series of exposures with known offsets between their
respective comb signals along the spectral direction of the CCD. In
this way we intended to more strongly constrain the IPS variations in
a single CCD row. However, a series of shorter exposures
\emph{without} offsets between them revealed short time-scale
($\la$1\,min) random shifts between exposures of up to
$\approx$0.4\,pixels (or $\approx$40\,\ms) -- see
Fig.~\ref{fig:shifts}. This precludes a thorough analysis of the
series of offset comb exposures for constraining IPS
variations. Nevertheless, we used simulated spectra whose parameters
closely matched our real spectra to demonstrate that, with a stable
spectrograph, such a series of comb exposures should allow symmetric
departures from a flat average IPS map of a single CCD row (320
pixels) to be strongly constrained -- see Fig.~\ref{fig:simIPS}.

Despite demonstrating that the \emph{symmetric} component of average
IPS variations can be constrained using LFCs, it is clear that,
without knowing \emph{a priori} where the comb modes fall with respect
to the CCD grid, any \emph{asymmetric} component of the IPS map cannot
be reliably measured. This could be an important aspect of IPS
variations to measure for some applications, especially those where
long-term, effectively absolute calibration is required (e.g.~the
real-time observation of the drift in Lyman-$\alpha$ forest lines in
quasar spectra). In future, perhaps the relationship between the
physical position of the comb modes and the CCD grid might be
established through some kind of physical CCD masking procedure, or by
`switching off' some CCD pixels and obtaining a series of offset comb
exposures. Exploring such possibilities may be required to fully
realise absolute wavelength calibration using CCDs in astronomy.

\section*{Acknowledgments}

We are indebted to the truly exceptional staff at the Australian
Astronomical Observatory for enabling the experiment reported here. In
particular, the expertise and resourcefulness of Stuart Barnes, Steve
Lee, Rob Dean, Stephen Marsden, Doug Gray, Steve Chapman, John Collins
and Winston Campbell ensured its success. MTM thanks the Australian
Research Council for a QEII Research Fellowship (DP0877998), while AL,
CL and PL thank it for supporting this work under DP0877938.


\bspsmall

\label{lastpage}

\end{document}